\documentclass[aps,twocolumn,pra]{revtex4}
\usepackage{epsfig}
\usepackage{graphicx}
\usepackage{amssymb,amsmath}
\usepackage{bbm}
\usepackage{moreverb}
\usepackage[dvips]{color}

\newcommand {\be}{\begin{equation}}
\newcommand {\ee}{\end{equation}}
\newcommand {\ba}{\begin{eqnarray}}
\newcommand {\ea}{\end{eqnarray}}

\begin{document}

\title{The Hyperfine Molecular Hubbard Hamiltonian}
\author{M. L. Wall and L. D. Carr}
\affiliation{Department of Physics, Colorado School of Mines, Golden, CO 80401, USA}

\begin{abstract}
An ultracold gas of heteronuclear alkali dimer molecules with hyperfine structure loaded into a one-dimensional optical lattice is investigated.  The \emph{Hyperfine Molecular Hubbard Hamiltonian} (HMHH), an effective low-energy lattice Hamiltonian, is derived from first principles.  The large permanent electric dipole moment of these molecules gives rise to long range dipole-dipole forces in a DC electric field and allows for transitions between rotational states in an AC microwave field.  Additionally, a strong magnetic field can be used to control the hyperfine degrees of freedom independently of the rotational degrees of freedom.  By tuning the angle between the DC electric and magnetic fields and the strength of the AC field it is possible to control the number of internal states involved in the dynamics as well as the degree of correlation between the spatial and internal degrees of freedom.  The HMHH's unique features have direct experimental consequences such as quantum dephasing, tunable complexity, and the dependence of the phase diagram on the molecular state.
\end{abstract}

\maketitle

\section{Introduction}
\label{sec:introduction}

Ultracold molecular gases are of interest in many subfields of science ranging from precision science to quantum simulation of many-body Hamiltonians~\cite{carr2009b}.  Recent success using the STIRAP (STImulated Raman Adiabatic Passage) method has allowed experimentalists to produce a gas of KRb molecules close to Fermi degeneracy, in the ground rovibrational state, and in a specific hyperfine level~\cite{niKK2008, Ospelkaus2009b}.  Rovibonic ground state molecules have also been formed for polar LiCs\cite{Deiglmayr2008b} as well as nonpolar Cs$_2$\cite{Danzl2010} and Rb$_2$\cite{Winkler2007}, with studies on other species currently underway\cite{Pilch2009, Voigt2009}.  To reach the quantum degenerate regime one must have all molecules in the same quantum state, a task which is complicated by the rich hyperfine structure of alkali dimer molecules.  Thus, a number of recent works~\cite{Hutson2009, Ran2009, ospelkaus2009} have investigated the single-molecule microwave spectra to find a route by which all molecules are transferred to the lowest hyperfine state, yielding a gas of \emph{absolute} ground state molecules.

From the condensed matter perspective, ultracold gases are enticing in their capacity to act as \emph{quantum simulators}~\cite{feynmanRP1982,lewensteinM2007}.  Such specialized quantum computers allow for the study of complex many-body Hamiltonians in a setting where many parameters are amenable to experimental control.  From this point of view, it is natural to ask how the various degrees of freedom in the quantum simulator may be controlled and used as resources.  Theoretical proposals for many-body physics using ultracold molecules have so far focused only on the rotational degrees of freedom in $^1\Sigma$ molecules with external fields~\cite{micheli2007, carr2009a} or on the hyperfine degree of freedom in $^2\Sigma$ molecules without external fields~\cite{Brennen2007}.  In this work we study $^1\Sigma$ molecules in strong fields including the effects of hyperfine structure and discuss how the hyperfine degrees of freedom may be controllably accessed and manipulated as a resource for generating complex quantum dynamics.

For $^1\Sigma$ molecules it has been shown that the interaction of the rotational degrees of freedom with external electric fields allows for the tuning of the strength and range of the two-molecule interaction potential~\cite{micheli2007}.  Many of these results also hold for molecules with hyperfine structure, as the rotational and nuclear spin degrees are only weakly coupled in strong fields.  In particular, the application of a DC field and an optical trapping potential gives rise to a purely repulsive dipole-dipole interaction between molecules in reduced geometries.  Also, it has been shown that the combination of a strong uniform magnetic field and a suitably chosen microwave field allows for transitions between particular hyperfine single-molecule states, and that this may be used to transfer a collection of molecules that have been cooled to the rovibrational ground state but an excited hyperfine state to their hyperfine ground state~\cite{Hutson2009, Ran2009,ospelkaus2009}.  This idea also works in reverse: one can select the states which are involved in many-body dynamics with the ground state by judicious choice of the field strengths and geometries.  The HMHH reflects this fact; not only the parameters of the Hamiltonian but also the dimensionality and character of the basis are suited to experimental control.

This article is organized as follows.  In Sec.~\ref{sec:hfMHH} we introduce the HMHH, define its parameters, and discuss its novel experimental consequences.  This section contains the main results of the paper.  In Sec.~\ref{sec:der} we derive the HMHH from first principles and state the key assumptions underlying its derivation.  Finally, in Sec.~\ref{sec:conclusion}, we conclude.  Some details concerning the single molecule physics are provided in the appendices in the interest of completeness.

%Some well-known results concerning the single-molecule spectra of $^{40}$K$^{87}$Rb, the most experimentally relevant species for which the HMHH applies, are given in Appendices \ref{sec:internal} and \ref{sec:fields} in the interest of completeness.  For convenience, all of the matrix elements of the single-molecule Hamiltonian are given in Appendix \ref{sec:matelem}.

\section{Statement of the Hamiltonian and Experimental Consequences}
\label{sec:hfMHH}
The Hyperfine Molecular Hubbard Hamiltonian is
\begin{align}
\nonumber\hat{H}&=\sum_{\sigma}\Delta_{\sigma}\sum_i\hat{n}_{i\sigma}-\sum_{\sigma}t_{\sigma}\sum_{\langle i,j\rangle}\left[\hat{a}_{i\sigma}^{\dagger}\hat{a}_{j\sigma}+\mbox{h.c.}\right]\\
\nonumber&+\frac{1}{2}\sum_{\sigma,\sigma'}U_{\sigma\sigma'}\sum_{\langle i,j\rangle}\hat{n}_{i\sigma}\hat{n}_{j\sigma'}\\
\label{eq:HMHH}&-\frac{1}{2}\sum_{\sigma\sigma'}d_{\sigma\sigma'}E_{\mathrm{AC}}\sum_i\left[\hat{a}_{i\sigma}^{\dagger}\hat{a}_{i\sigma'}+\mbox{h.c.}\right]\, ,
\end{align}
where $\hat{a}_{i\sigma}$ destroys a bosonic or fermionic molecule in state $|\sigma\rangle$ on the $i^{th}$ lattice site, and the bracket notation $\langle\dots\rangle$ denotes that the sum is taken over nearest neighbors.  The single-molecule basis $\left\{ |\sigma\rangle\right\}$ takes into account the hyperfine interactions (Appendix \ref{sec:internal}) and static fields (Appendix \ref{sec:fields}) and the quantum number $\sigma$ is a composite index referring to both rotational and nuclear spin degrees of freedom.  The properties and dimensionality of this basis can be modified by the geometry and strength of the external fields, as will be discussed in more detail below.

The first term in the HMHH represents the energy offset of a molecule in state $|\sigma\rangle$ from a reference ground state.  The second term describes the tunneling of molecules between lattice sites and depends on the rotational state.  The third term describes resonant dipole-dipole interactions between molecules on neighboring sites.  The final term corresponds to transitions driven between states $|\sigma\rangle$ and $|\sigma'\rangle$ by an AC microwave field.  Here the transition dipole moment between two states $|\sigma\rangle$, $|\sigma'\rangle$ is $d_{\sigma\sigma'}\equiv\langle\sigma|\hat{d}_1|\sigma'\rangle$, where $\hat{d}_1\equiv \hat{\mathbf{d}}\cdot\mathbf{e}_1$ is the projection of the dipole operator along the space-fixed spherical basis direction $\mathbf{e}_1=-\left(\mathbf{e}_x+i\mathbf{e}_y\right)/\sqrt{2}$.

For $^{40}$K$^{87}$Rb, which is the most experimentally relevant species, the energy scales of the various terms are summarized in Table \ref{table:scales}.  The detunings $\Delta_{\sigma}$ are determined chiefly by the linear Zeeman effect, and so are tunable by the DC magnetic field, and will be similar for other molecular species.  The tunneling energy scale $t_\sigma$ is set by the recoil energy, and so will be similar for other alkali dimers.  The dipole-dipole energy scale $U_{\sigma\sigma'}$ is fixed by the permanent dipole moment, and so will change with the molecular species.  For example, LiCs has a dipole moment roughly 10 times larger than that of KRb, and so $U_{\sigma\sigma'}$ will be of order $25$kHz.  The scale of the AC term is determined by the power of the microwave field $E_{\mathrm{AC}}$, which is readily tunable.  The range of energies we have quoted represents the most interesting regime where the basic assumptions of our derivation hold.
\begin{table}[t]
\begin{tabular}{|c|c|}
\hline Term&Energy scale\\
\hline $\Delta_{\sigma}$&$\sim 1-100$kHz (depends on static field strengths)\\
\hline $t_{\sigma}$&$\sim 1$ kHz\\
\hline $U_{\sigma\sigma'}$&$\sim$250 Hz\\
\hline $d_{\sigma\sigma'}E_{\mathrm{AC}}$&$\sim 1-50$kHz \\
\hline
\end{tabular}
  \caption{Table of energy scales of the Hyperfine Molecular Hubbard Hamiltonian.  From top to bottom: energy $\Delta_{\sigma}$ of internal state $|\sigma\rangle$, relative to the ground state; tunneling $t_{\sigma}$; dipole-dipole interaction $U_{\sigma\sigma'}$; transition dipole moment $d_{\sigma\sigma'}$ due to the AC electric drive $E_{\mathrm{AC}}$.}
  \label{table:scales}
\end{table}

In the following sections we will justify the HMHH and list the essential assumptions underlying its derivation, but we first pause to note some of its novel properties.

\subsection{Quantum Dephasing}
The first property, which we call \emph{quantum dephasing}, was investigated previously for a molecular Hubbard Hamiltonian involving only rotational degrees of freedom~\cite{carr2009a}.  The effect, which is purely many-body in nature, may be summarized in this context as the destruction of coherent Rabi flopping due to the population of many spatial degrees of freedom in a many-body system driven at a single-molecule resonance.  This effect is also of interest in the more general context of oscillations in a many-body system that are damped by some intrinsic mechanism following a quench~\cite{huber2009, Barmettler2009}.

\begin{figure}[t]
\begin{center}
\epsfxsize=8cm \epsfysize=5.51 cm \leavevmode \epsfbox{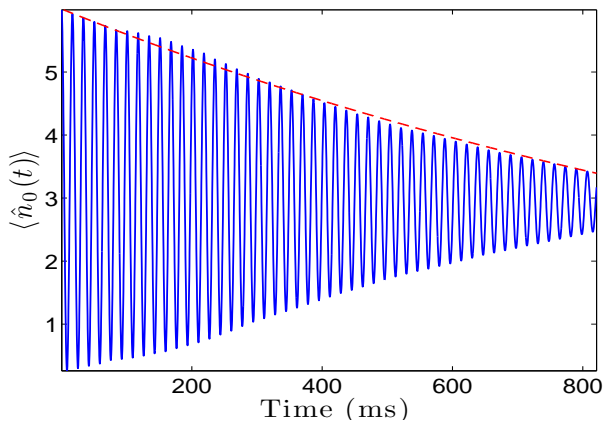}
\caption{\label{fig:deph} (Color online) \emph{Quantum dephasing in the HMHH.}  The plot shows the behavior of the total number in state $0$: $\langle \hat{n}_{0}\rangle\equiv \langle\sum_{i}\hat{n}_{i0}\rangle$ when the system evolves under the Hamiltonian \eqref{eq:HMHH}.  Quantum dephasing produces an emergent exponential envelope on the Rabi oscillation pattern between states 0 and 1.  Only the number of state 0 is shown for clarity.  The dashed red curve is an exponential envelope fit to $N\exp\left(-t/\tau\right)$ with $\tau=1441.17$ms.  The nonexponential behavior near $t=200$ is due to the finite size of the lattice.}
\end{center}
\end{figure}

Dephasing is strongest when the Rabi frequency is on the order of the tunneling energies and the difference in tunneling energies for the two internal modes is also comparable to these two scales.  For a system with two single-particle levels $0$ and $1$ and tunneling energies $t_0$ and $t_1$, respectively, this gives the condition $\Omega\sim t_0\sim t_1\sim \left|t_0-t_1\right|$, which can be achieved with the HMHH for reasonable parameter values.  The Rabi oscillations between the two internal states connected by the single molecule resonance damp out exponentially in time with an emergent time scale $\tau$ which can be measured experimentally, see Fig.~\ref{fig:deph}.  Dephasing can be observed in the structure factors
\begin{align}
S_{\pi}^{\sigma\sigma'}&=\frac{1}{L}\sum_{i,j=1}^{L}\left(-1\right)^{i-j}\langle \hat{n}_{i\sigma}\hat{n}_{j\sigma'}\rangle\, ,
\end{align}
where $L$ is the number of lattice sites; $S_{\pi}^{\sigma\sigma'}$ can be measured in scattering experiments~\cite{A&M}.

\subsection{Internal State Dependence of Phase Diagram}

\begin{figure}[t]
\begin{center}
\epsfxsize=8cm \epsfysize=6.2 cm \leavevmode \epsfbox{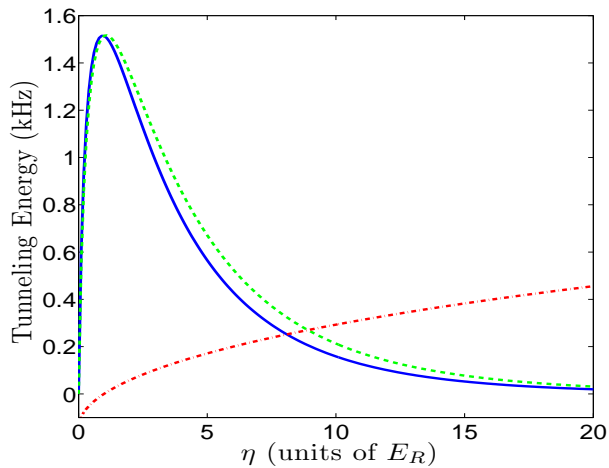}
\caption{  \label{fig:tunn}  (Color online) \emph{Tunneling matrix elements in a DC electric field.} Tunneling energies (in kHz) of the $N=0$ (solid blue line) and $N=1$ (dashed green line) rotational states and their difference divided by their arithmetic mean, $2\left(t_1-t_0\right)/\left(t_1+t_0\right)$, (dash-dotted red line) for KRb in a field of 10kV/cm as a function of the effective isotropic lattice height $\eta\equiv \bar{\alpha}\left|\mathbf{E}_{\mathrm{opt}}\right|^2$ (in recoil energy units).  The values of the polarizability tensor are taken from Ref.~\cite{deiglmayr2008}.}
\end{center}
\end{figure}

The dependence of the tunneling energy $t_{\sigma}$ on the internal state $\sigma$ makes the borders of the phase diagram shift strongly (e.g. by a factor of 2).  This dependence is shown explicitly in Fig.~\ref{fig:tunn}.  Thus, by preparing a collection of molecules in multiple internal states one can study interactions of many-body systems in different quantum phases and possibly far from equilibrium.  Possibilities for quantum statics include studying the properties of phase equilibria as a function of population imbalance and effective mass (as determined by the tunneling energy)~\cite{iskin2007b}.  Also, as the difference in tunneling energy between different modes depends only on the elements of the molecular polarizability tensor, measuring the borders of the static phase diagram for different internal states also provides a novel means to measure this tensor.  Possibilities for quantum dynamics include the study of quench phenomena for interacting many-body systems in different quantum phases.

\subsection{Tunable Complexity}

\label{sec:der}
\begin{figure}[t]
\begin{center}
\epsfxsize=9cm \epsfysize=6.2 cm \leavevmode \epsfbox{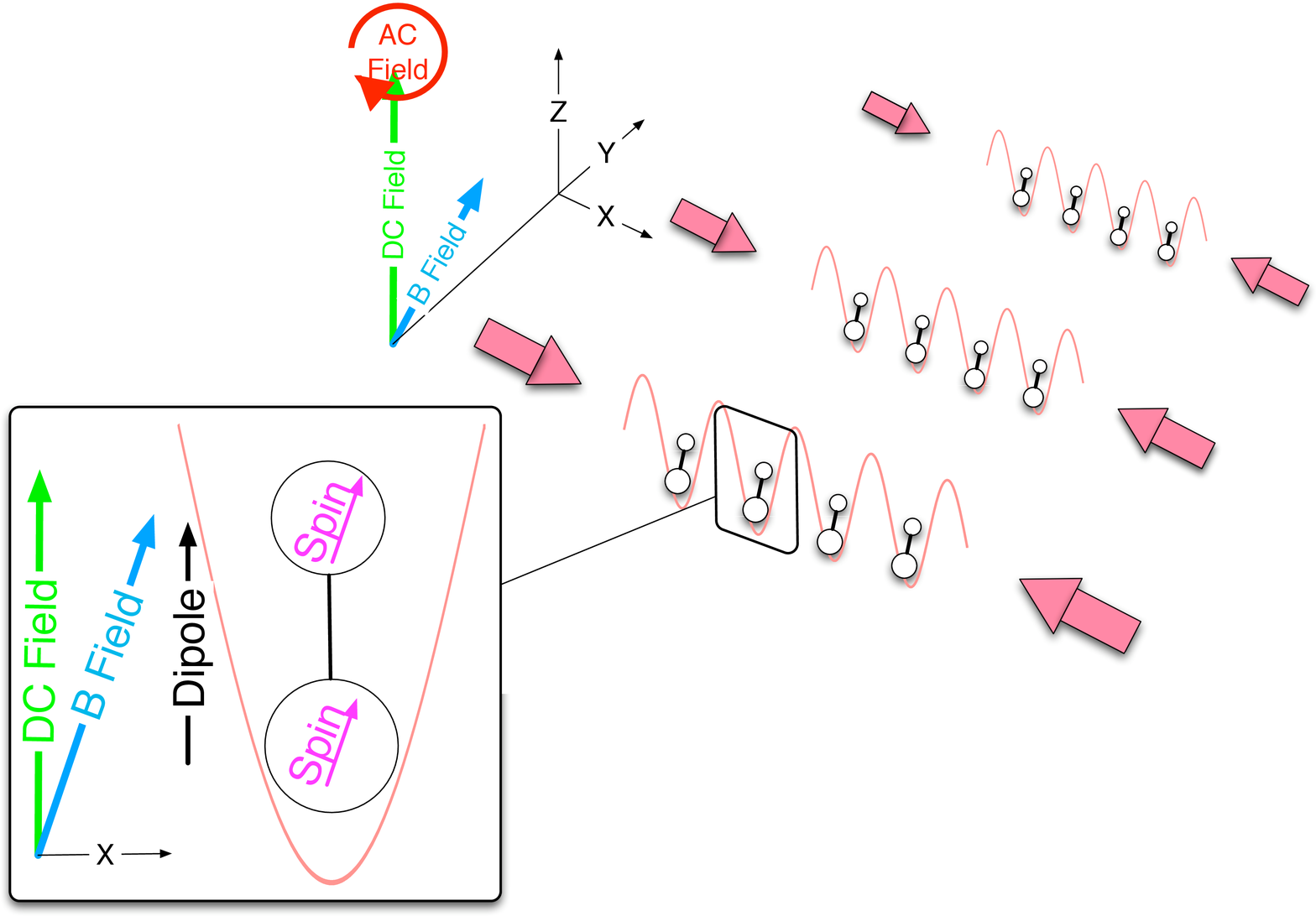}
\caption{\label{fig:exSch}  (Color online) \emph{Geometry of the HMHH.}  Counter-propagating laser beams along the $y$ and $z$ directions create an array of 1D tubes, and an additional pair of laser beams along $x$ creates a lattice potential.  A strong DC field orients the dipoles along the direction perpendicular to motion, and a magnetic field orients the nuclear spins.  An AC field of circular space-fixed polarization drives transitions between internal levels.}
\end{center}
\end{figure}

A final noteworthy property which was not present in the molecular Hubbard Hamiltonians previously studied is the possibility of \emph{tunable complexity}.  By complexity we mean that the system is comprised of many interacting degrees of freedom and displays emergent behavior such as the dephasing discussed above.  Tunability refers to the fact that we may alter the \emph{number} of internal degrees of freedom that are accessed dynamically as well as the \emph{timescale} of their relative interactions.  The key point for tunability is that the electric and magnetic fields affect different degrees of freedom: the electric dipole moment and nuclear spins, respectively.  We illustrate this concept, and the corresponding geometries and polarizations needed for experiments, in Fig.~\ref{fig:exSch}.

In slightly more detail, tunability is achieved as follows. In the presence of an electric field aligned along the $z$ direction, dipole moments are induced between states having the same nuclear spin projection along the field.  The introduction of a strong magnetic field defines an effective axis of quantization for the nuclear spins while leaving the rotational structure unchanged because of the strong nuclear Zeeman effect, the weak rotational Zeeman effect, and the presence of only weak (quadrupole) coupling between the rotational and nuclear spin degrees of freedom.  In the presence of a strong magnetic field that is not collinear with the electric field it is therefore possible to induce dipole moments between states with different hyperfine quantum numbers.

Thus, by changing the relative angle between the electric and magnetic fields one can control the number of states accessible from a particular state.  The power of the applied AC field determines the interaction scale and the Rabi frequency of these dipole couplings, and the strength of the magnetic field determines the energetic splittings between states, in turn determining the relative rates of internal state population.  The HMHH may therefore be used as a quantum simulator of a \emph{quantum complex system} where the number and timescale of the internal components may be dynamically altered.  Precise measures of complexity and simulations displaying characteristic behavior in various regimes will be discussed in future work~\cite{wallforthcoming}.

\section{Derivation of the Hyperfine Molecular Hubbard Hamiltonian}
We consider the experimental setup shown schematically in Fig.~\ref{fig:exSch}.  Counter-propagating laser beams along the ${y}$ and ${z}$ directions create a series of 1D optical lattice ``tubes.''  The intensity of these beams is such that the tubes are isolated from one another, and the lattice spacing is chosen (e.g. by crossed beams) such that the dipole-dipole interaction along $y$ and $z$ is negligible on experimental timescales.  An additional pair of beams creates a lattice potential along the $x$-direction.  The experimental techniques required to create this setup have been well established for ultracold atoms~\cite{paredes2004,tolra2004,kinoshita2006}.  In addition to the lattice potential there is a uniform DC electric field along the ${z}$ direction, a uniform magnetic field which lies in the $xz$ plane, and an AC microwave field propagating in $z$ which is assumed to have circular polarization $q=1$ in the space-fixed spherical basis.

In the lattice is an ultracold quantum degenerate gas of $^1\Sigma$ heteronuclear molecules characterized by permanent electric dipole moment $d$, rotational constant $B_N$, rotational angular momentum $\mathbf{N}$~\footnote{We reserve $\mathbf{J}$ for future studies involving nonzero orbital or electronic spin angular momentum.}, and nuclear spins $\mathbf{I}_1$ and $\mathbf{I}_2$. Both nuclear spins are taken to be greater than one-half, so that both nuclei have nonzero electric quadrupole moments.  In second quantization the full low-energy Hamiltonian for this setup is 
\begin{eqnarray}
\label{eq:fullHami} \hat{H}&=&\int \!\! d\mathbf{r}\,\hat{\psi}^{\dagger}\!\left(\mathbf{r}\right)\left[\hat{H}_{\mathrm{in}}+\hat{H}_{\mathrm{F}}+\hat{H}_{\mathrm{AC}}+\hat{H}_{\mathrm{kin}}+\hat{H}_{\mathrm{opt}}\right]\hat{\psi}\!\left(\mathbf{r}\right)\nonumber\\
&&\!\!\!\!+\frac{1}{2}\int\!\! d\mathbf{r}d\mathbf{r}'\,\hat{\psi}^{\dagger}\!\left(\mathbf{r}\right)\hat{\psi}^{\dagger}\!\left(\mathbf{r}'\right)\hat{H}_{\mathrm{DD}}\!\left(\left|\mathbf{r}-\mathbf{r}'\right|\right)\hat{\psi}\!\left(\mathbf{r}'\right)\hat{\psi}\!\left(\mathbf{r}\right)\!,
\end{eqnarray}
where
\begin{align}
\label{eq:int}\hat{H}_{\mathrm{in}}=&\hat{H}_{\mathrm{rot}}+\hat{H}_{\mathrm{scal}}+\hat{H}_{\mathrm{tens}}+\hat{H}_{\mathrm{r-s}}+\hat{H}_{\mathrm{quad}}\\
\nonumber=&B_N\hat{\mathbf{N}}^2+c_4\hat{\mathbf{I}}_1\cdot\hat{\mathbf{I}}_2+c_3\hat{\mathbf{I}}_1\cdot\tilde{T}\cdot\hat{\mathbf{I}}_2+\sum_{i=1}^{2}c_i\hat{\mathbf{N}}\cdot\hat{\mathbf{I}}_i\\
\nonumber&+\sum_{i=1}^{2}\hat{\tilde{\mathbf{V}}}_i\cdot\hat{\tilde{\mathbf{Q}}}_i\, ,\\
\hat{H}_{\mathrm{F}}=&-g_r\mu_N\hat{\mathbf{N}}\cdot\mathbf{B}-\sum_{i=1}^{2}g_i\mu_N\left(1-\sigma_i\right)\hat{\mathbf{I}}_i\cdot\mathbf{B}\\
\nonumber&-\mathbf{E}_{\mathrm{DC}}\cdot\hat{\mathbf{d}}\, ,\\
\hat{H}_{\mathrm{AC}}=&-\mathbf{E}_{\mathrm{AC}}\cdot\hat{\mathbf{d}}\, ,\\
\hat{H}_{\mathrm{kin}}=&\frac{\hat{\mathbf{p}}^2}{2m}\, ,\\
\hat{H}_{\mathrm{opt}}=&-\mathbf{E}_{\mathrm{opt}}^{\star}\cdot\hat{\tilde{\alpha}}\left(\omega_{\mathrm{opt}}\right)\cdot\mathbf{E}_{\mathrm{opt}}\, ,\\
\hat{H}_{\mathrm{DD}}\left(R\right)=&\frac{\hat{\mathbf{d}}_1\cdot \hat{\mathbf{d}}_2-3\left(\hat{\mathbf{d}}_1\cdot\mathbf{e}_R\right)\left(\mathbf{e}_R\cdot\hat{\mathbf{d}}_2\right)}{R^3}\, .
\end{align}
The first line of Eq.~\eqref{eq:fullHami} is comprised of single-molecule terms.  In order, these are $\hat{H}_{\mathrm{in}}$, the Hamiltonian governing the internal rotational and nuclear spin degrees of freedom; $\hat{H}_{\mathrm{F}}$, the interaction of the molecule with externally applied DC electric and magnetic fields; $\hat{H}_{\mathrm{AC}}$, the interaction of the molecule with an AC microwave field; $\hat{H}_{\mathrm{kin}}$, the kinetic energy of the molecule; and  $\hat{H}_{\mathrm{opt}}$, the interaction of the molecule with the optical lattice potential.  The second line of Eq.~\eqref{eq:fullHami} is the two-molecule resonant dipole-dipole force.  The main assumptions underlying this Hamiltonian and our subsequent analysis are the following.

First, the individual molecules are assumed to be in their electronic and vibrational ground states, and it is assumed that none of these degrees of freedom can be excited at the large intermolecular separations and low temperatures/relative energies that we consider.

Second, the characteristic trapping potential length is chosen large enough compared to the internuclear axis to assume spherical symmetry, i.e.~a locally constant potential.

Third, we consider only the lowest two rotational levels.  All AC fields will be sufficiently weak to allow this assumption.  We also work in the rotating wave approximation, which requires that the detuning be small compared to the driving frequency.

Fourth, we consider all molecules to be in the lowest Bloch band.  The AC Rabi frequencies are chosen to be small ($\sim$1-50kHz) in comparison with the lattice bandwidth ($\sim$10$E_R\sim$100kHz) to ensure this assumption.

Fifth, we work in the ``hard-core" limit where at most one molecule is allowed per site.  This is enforced by strongly repulsive dipole-dipole interactions on-site, caused by our $z$-alignment of the electric field, as sketched in Fig~\ref{fig:exSch}.  We consider the lattice spacing large enough to include only nearest-neighbor dipole-dipole interactions.  We neglect the effects of chemical reactions or hyperfine changing collisions which occur at very short range.

Sixth, we neglect dipole-dipole interactions between molecules in different 1D ``tubes.''  For a consistent level of approximation this requires the tubes to be separated by twice the lattice spacing.  This can be achieved in principle using crossed beams to create larger lattice spacings.

Seventh, we consider only pairwise interactions of the molecules, neglecting three and higher-body interactions.  This is valid for KRb because the permanent dipole moment $d=0.566$D is rather small.  For molecules such as LiCs with larger permanent dipole moments, the three-body interaction can play a significant role~\cite{buechler2007}.

To derive a Hamiltonian of Hubbard type from Eq.~\eqref{eq:fullHami} we follow the standard prescription~\cite{lewensteinM2007} of expanding the field operators of our second-quantized Hamiltonian in a Wannier basis of single-molecule states centered at a particular discrete position $\mathbf{r}_i$:
\begin{eqnarray}
\hat{\psi}&=\sum_{i}\sum_{\sigma} \hat{a}_{i\sigma}w_{\sigma}\left(\mathbf{r}-\mathbf{r}_i\right)\,,
\end{eqnarray}
where $i$ is a site index and $\sigma$ an index denoting the internal state of the molecule.  The Wannier basis we use is the basis which diagonalizes the internal plus static field Hamiltonians $\hat{H}_{\mathrm{in}}+\hat{H}_{\mathrm{F}}$ and in which all states with $N=1$ rotate with frequency $\omega$, where $\omega$ is the frequency of the applied AC electric field.  With the field operator written in this manner, we find the Hubbard parameters 
\begin{align}
 t_{\sigma}&\equiv-\int\!\! d\mathbf{r}\,w^{\star}_{\sigma}\left(\mathbf{r}-\mathbf{r}_i\right)\left[\hat{H}_{\mathrm{kin}}+\hat{H}_{\mathrm{opt}}\right]w_{\sigma}\left(\mathbf{r}-\mathbf{r}_{i+1}\right)\, ,\\
\Delta_{\sigma}&\equiv \int\!\! d\mathbf{r}\,w^{\star}_{\sigma}\left(\mathbf{r}-\mathbf{r}_i\right)\left[\hat{H}_{\mathrm{in}}+\hat{H}_{\mathrm{F}}\right]w_{\sigma}\left(\mathbf{r}-\mathbf{r}_i\right)\, ,
\end{align}
and
\begin{align}
-d_{\sigma\sigma'}E_{\mathrm{AC}}&\equiv \int\!\! d\mathbf{r}\,w^{\star}_{\sigma}\left(\mathbf{r}-\mathbf{r}_i\right)\hat{H}_{\mathrm{AC}}w_{\sigma'}\left(\mathbf{r}-\mathbf{r}_i\right)\, ,\\
U_{\sigma\sigma'}&\equiv \int\!\! d\mathbf{r} d\mathbf{r}'\,w^{\star}_{\sigma}\left(\mathbf{r}-\mathbf{r}_i\right)w^{\star}_{\sigma'}\left(\mathbf{r}'-\mathbf{r}_{i+1}\right)\\
\nonumber &\times H_{\mathbf{DD}}\left(\mathbf{r}-\mathbf{r}'\right)w_{\sigma}\left(\mathbf{r}-\mathbf{r}_i\right)w_{\sigma'}\left(\mathbf{r}'-\mathbf{r}_{i+1}\right)\, .
\end{align}

The detunings $\Delta_{\sigma}$ are determined by the single-molecule spectra, which are well-known~\cite{Hutson2009,Hutson2008}.  In the interest of the present article's completeness, we have included appendices reviewing the basic results and explaining them in the context of the present problem.  In the following sections we discuss the remaining Hubbard parameters.

\subsection{Tunneling Energies}
A key component of the realization of many-body Hamiltonians using ultracold molecules is the presence of a far off-resonant optical lattice which confines the molecules in a reduced geometry.  The Hamiltonian of this interaction is
\begin{align}
 \hat{H}_{\mathrm{opt}}&=-\mathbf{E}_{\mathrm{opt}}^{\star}\left(\mathbf{r},\omega_{\mathrm{opt}}\right)\cdot\hat{\tilde{\alpha}}\left(\omega_{\mathrm{opt}}\right)\cdot\mathbf{E}_{\mathrm{opt}}\left(\mathbf{r},\omega_{\mathrm{opt}}\right)\, ,
\end{align}
where $\mathbf{E}_{\mathrm{opt}}\left(\mathbf{r},\omega_{\mathrm{opt}}\right)$ is the optical lattice field and $\hat{\tilde{\alpha}}\left(\omega_{\mathrm{opt}}\right)$ is the polarizability tensor operator of the molecule, evaluated at the optical lattice frequency $\omega_{\mathrm{opt}}$.  In our notation, the circumflex accent (the `hat') denotes an operator, the tilde denotes a rank 2 tensor, and boldface denotes a rank 1 tensor, or vector.  This optical potential couples to the electronic degrees of freedom and is detuned from resonance by an amount several orders of magnitude larger than any hyperfine splittings.  Thus dependence on the hyperfine quantum numbers in negligible.  For tight optical traps, the optical trap potential at each well is close to that of a harmonic trap plus a small state-dependent tensor shift of the trap frequency affecting levels with $N>0$ due to the polarizability anisotropy~\cite{carr2009a}.

When the optical potential is combined with the kinetic portion of the Hamiltonian and evaluated in the Wannier basis one obtains the tunneling energies. As the tunneling energies are independent of the hyperfine quantum numbers, we can use results obtained in the case of only rotational degrees of freedom, derived in our earlier work~\cite{carr2009a}.  Then the tunneling energies in the eigenbasis of $\hat{H}_{\mathrm{rot}}$, $|NM_N\rangle$, are given by
\begin{align}
\frac{\tilde{t}_{NM_N}}{E_R}&=A\left(\frac{V_{NM_N}}{E_R}\right)^B\exp\left(-C\sqrt{\frac{V_{NM_N}}{E_R}}\right)
\end{align}
where $A=1.397$, $B=1.051$, and $C=2.121$ are fit parameters~\cite{reyAM2004}, $E_R$ the recoil energy, and
\begin{align}
V_{NM_N}&=\left|\mathbf{E}_{\mathrm{opt}}\right|^2\left[\bar{\alpha}+\frac{2\Delta\alpha}{3}\frac{N\left(N+1\right)-3M_N^2}{\left(2N-1\right)\left(2N+3\right)}\right]\,
\end{align}
is the effective lattice height for the $|NM_N\rangle$ level.  Here $\bar{\alpha}$ is the average polarizability and $\Delta\alpha$ the polarizability anisotropy.

In the presence of a DC field the rotational levels become mixed, leading to new effective tunneling energies which we denote as ${t}_{NM_N}$, with $N$ and $M_N$ the corresponding zero field values.  This hybridization of rotational levels in principle also allows tunneling events which change the rotational state of the molecule, but we can ignore such events because the rotational level separation is much larger than the tunneling energies.  The effective tunneling for the $N=0$ and $N=1$, $M_N=\pm1$ levels is shown in Fig.~\ref{fig:tunn}.  The scale is set by the recoil energy, which is $2\pi\times 1.44\, $kHz for KRb in a 1054 nm optical lattice.

\subsection{Two-Molecule Interactions}
\label{sec:twomolecule}
Heteronuclear $^1\Sigma$ molecules posses permanent dipole moments, and thus interact via a dipole-dipole interaction
\begin{align}
\hat{H}_{\mathrm{DD}}\left(\mathbf{R}\right)&=\frac{\hat{\mathbf{d}}_1\cdot \hat{\mathbf{d}}_2-3\left(\hat{\mathbf{d}}_1\cdot\mathbf{e}_R\right)\left(\mathbf{e}_R\cdot\hat{\mathbf{d}}_2\right)}{R^3}\, ,
\end{align}
where $\mathbf{R}\equiv \mathbf{r}_2-\mathbf{r}_1$, $\mathbf{e}_R$ is a unit vector in the direction of $\mathbf{R}$, and $\hat{\mathbf{d}}_i$ is the vector dipole operator of the $i^{th}$ molecule.  In the absence of external fields, this interaction is off-resonant, leading to a van der Waals interaction $\hat{H}_{\mathrm{DD}}\left(\mathbf{R}\right)\sim R^{-6}$, but in the presence of electric fields resonant dipoles are induced and the interaction displays a resonant $R^{-3}$ behavior in addition to the $R^{-6}$ behavior.

The anisotropic nature of the dipole-dipole force has been experimentally shown to dominate the rethermalization behavior of a molecular gas via inelastic collisions~\cite{niKK2010}.  This is because a ``head-to-tail'' arrangement of molecules leads to an attractive potential, whereas ``side-to-side'' interactions are repulsive.  To ensure the stability of an ultracold molecular ensemble and to prevent losses from inelastic collisions it is crucial therefore not only to orient the dipoles using a DC field, but also to confine the molecules in a reduced geometry.  A thorough discussion of the nature of the two-molecule spectra for $^1\Sigma$ molecules without hyperfine structure and its implications for stability in two dimensions is presented in Ref.~\cite{micheli2007}.  Diagonalization of the full two-molecule Hamiltonian is impractical when hyperfine structure is included due to the very large matrices that result.  Instead, we argue based on comparisons of length and energy scales that the hyperfine structure is negligible during the collisional processes which occur in our proposed setup.

Our reduced geometry is imposed by the optical lattice described earlier.  Namely, we consider the case where the molecules are confined to move only along the $x$ direction and a DC field polarized along the $z$ direction orients the dipoles such that all collisions are side-to-side and repulsive.  The dipole-dipole interaction in this geometry reduces to
\begin{align}
\nonumber \hat{H}_{\mathrm{DD}}=&\frac{1}{R^3}\Big[\hat{d}_0\otimes \hat{d}_0+\frac{1}{2}\left(\hat{d}_{-1}\otimes \hat{d}_1+\hat{d}_1\otimes \hat{d}_{-1}\right)\\
&-3\left(\hat{d}_{-1}\otimes \hat{d}_{-1}+\hat{d}_1\otimes \hat{d}_1\right)\Big]\, ,
\end{align}
where $\hat{d}_q\equiv \hat{\mathbf{d}}\cdot \mathbf{e}_q$ is the component of the dipole operator along the $q$ direction in the space-fixed spherical basis.  For $z$-polarized electric field, the only diagonal components are those involving $d_0$.  The components of the interaction involving $d_{\pm 1}$ couple states with $\Delta M_N=\pm 1$ that are separated in energy by an amount of order the rotational constant for the DC fields we consider (see Fig.~\ref{fig:GHZStark}). Contributions from these components are suppressed at distances greater than $r_B\equiv \left(d^2/B\right)^{1/3}$, of order a few nanometers.  Thus, at the nearest-neighbor distance in a 1054nm optical lattice we consider only the diagonal elements of the dipole-dipole interaction.  This restriction gives rise to the two-body term
\begin{align}
 \hat{H}_{\mathrm{DD}}&=\frac{1}{2}\sum_{\sigma\sigma'}U_{\sigma\sigma'}\sum_{\langle i,j\rangle}\hat{n}_{i\sigma}\hat{n}_{j\sigma'}\,,
\end{align}
where
\begin{align}
 U_{\sigma\sigma'}&=\frac{d_{\sigma}d_{\sigma'}}{\left(\lambda/2\right)^3}\, .
 \label{eqn:easyU}
\end{align}
In Eq.~(\ref{eqn:easyU}) $d_{\sigma}$ is the resonant dipole moment of state $|\sigma\rangle $ and $\lambda$ is the wavelength of the optical lattice.  We assume that the long-range repulsive diagonal $d_0$ portion of the dipole-dipole interaction is strong enough to prevent both the occupation of any one lattice site by more than one molecule and access to the region where hyperfine-changing collisions involving the $d_{\pm1}$ dipole moments occur.

\subsection{Interactions with static external fields}
\label{sec:crossedfields}
The spectral properties of $^1\Sigma$ molecules in co-linear DC electric and magnetic fields have been elucidated elsewhere in the literature~\cite{Hutson2008,Hutson2009,Ran2009}, and the basic results of the analysis are given in Appendix \ref{sec:fields} for the reader's convenience.  In this section, we focus on the properties of such molecules in \emph{non-co-linear} fields, in particular on the dipole moments.

The behavior of the molecular dipole moments are controlled by an external DC electric field which mixes rotational levels of opposite parity and thus orients the molecule.  However, a DC field does not couple to the nuclear spins. So for a $z$-polarized field the selection rules $\Delta M_1=0$, $\Delta M_2=0$ are enforced, where $M_1$ and $M_2$ are the nuclear spin projections along the field direction.  In contrast, a magnetic field couples strongly to the nuclear spins but only weakly to the rotational angular momentum due to the relative sizes of the $g$-factors~\cite{Hutson2008}.  The magnetic field Hamiltonian thus has eigenstates which are energetically distinct nuclear spin states with a quantization axis given by the field direction.  It is in this sense that we say the magnetic field defines an effective axis of quantization for the nuclear spins.  Thus, in the absence of internal couplings of the rotational and hyperfine degrees of freedom they may be manipulated independently: the rotational angular momentum with an electric field and the nuclear spin angular momenta with a magnetic field.

\begin{figure}[t]
\begin{center}
\epsfxsize=9cm \epsfysize=6.2 cm \leavevmode \epsfbox{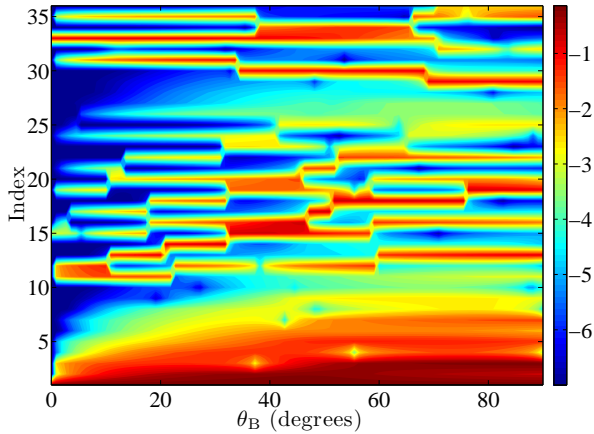}
\caption{\label{fig:dipsmear} (Color online) \emph{Distribution of dipolar character}.  The colorbar shows the logarithm of the transition dipole moment with the ground state, $\log\langle \mathrm{g.s.}|\hat{d}_1|i\rangle$, as a function of the angle between the magnetic and electric fields $\theta_B$ and the state index(ordered by energy).  Changing the angle between the electric and magnetic fields breaks the nuclear spin projection selection rule and allows for transition dipole moments between many states.  Only dipole moments greater than $10^{-7}$ are displayed. }
\end{center}
\end{figure}

The presence of nuclear quadrupole couplings in alkali dimer molecules couples states with the same total angular momentum projection $M_F$ but different rotational and nuclear spin projections.  For example, in KRb, the interaction couples $|\sigma\rangle = |N=1, M_N=0, M_{\mathrm{K}}, M_{\mathrm{Rb}} \pm 1\rangle$ to $|\sigma'\rangle = |N=1, M_N=\pm1, M_{\mathrm{K}}, M_{\mathrm{Rb}} \rangle$ with the latter accounting for $\sim 10\%$ of the state in the absence of fields.  (The interaction also couples the $M_{\mathrm{K}}\pm 1$ states, but the coupling constant $\left(eqQ\right)_{\mathrm{K}}$ is significantly smaller than $\left(eqQ\right)_{\mathrm{Rb}}$ and so the mixing is negligible in comparison)~\cite{ospelkaus2009}.  Clearly, since the $N=0$ state has only one projection $M_N=0$, the nuclear quadrupole interaction leaves this level unaffected.  In a DC electric field where the rotational levels become mixed, the states correlating with the $N=0$ levels and $N=1$ levels both display quadrupole effects, but these effects are still not identical.  In strong fields the Zeeman effect dominates over the quadrupole coupling, allowing control over the nuclear spins that displays  a weak dependence on the rotational level.

Thus, a strong magnetic field defines an effective axis of quantization for the nuclear spins, resulting in nuclear spin states which are superpositions of states in the basis with the axis of quantization along the electric field axis.  This implies that by changing the angle of the magnetic field with respect to the electric field, it is possible to change the number of states which are coupled by transition dipole moments.  This is illustrated in Fig.~\ref{fig:dipsmear}, which shows the logarithm of the transition dipole moment with the ground state as a function of the angle between the DC and magnetic fields $\theta_B$ and a state index (ordered by energy).  The lowest state index denotes the lowest energy state in the $N=1$ manifold.  When the fields are co-linear, one state dominates the dipole spectrum.  As the angle changes the dipolar character becomes spread over many states.  These transition dipole moments allow the states to couple in an AC microwave field and generate complex dynamics.

\subsection{Interaction with an AC microwave field}

The introduction of an AC microwave field contributes to the Hamiltonian in a similar way to a DC field.  In addition, the inherent time dependence allows for circular and linear polarization as well as the possibility of driving transitions between internal states.  In the absence of hyperfine structure, an AC field of spherical polarization $q$ couples the $|N=0,M_N=0\rangle$ and $|N=1,M_N=q\rangle$ levels, leading to an effective two-level system in the Floquet picture~\cite{micheli2007}.  In the presence of hyperfine structure, states with different total angular momentum projections $M_F$ in the $N\ge 1$ manifolds become mixed due to the electric quadrupole interaction.  Thus no rigorous selection rules can be established.  This complicates the issue of addressing single hyperfine states using microwave fields, but it also allows the hyperfine state to be changed using microwave fields.  Addressing a single hyperfine state can be achieved by the application of a strong magnetic field such as those used in the STIRAP procedure, which defines the projections sufficiently to suppress transitions to non-target hyperfine states~\cite{Hutson2009}.  In the presence of an electric field, this last comment holds only in the case where the two fields are co-linear.  When the fields are not co-linear many states can be accessed from any one state via a microwave transition due to the behavior of the transition dipole moments in crossed fields, as was described in Sec.~\ref{sec:crossedfields}.

We choose the polarization of the AC field to be purely circular, $q_{\mathrm{AC}}=1$. A component along $q=0$ would lead to rapid oscillation of the eigenenergies because the $d_0$ moments induced by the electric field couple to the AC field, and this complicates the analysis.  Furthermore, we consider Rabi frequencies which are much less than the bandwidth of the optical lattice so that our approximation of being in the lowest Bloch band remains valid and we are also justified in using a rotating wave approximation.  The above considerations together with the single-molecule AC Hamiltonian
\begin{align}
\hat{H}_{\mathrm{AC}}&=-\hat{\mathbf{d}}\cdot\mathbf{E}_{\mathrm{AC}}=-\hat{d}_qE_{\mathrm{AC}}e^{-i\omega t}+\mathrm{h.c.}
\end{align}
lead directly to the second quantized Hamiltonian
\begin{align}
\hat{H}_{\mathrm{AC}}&=-\frac{1}{2}\sum_{\sigma\sigma'} d_{\sigma\sigma'}E_{\mathrm{AC}}\sum_i\left[\hat{a}_{i\sigma}^{\dagger}\hat{a}_{i\sigma'}e^{i\omega t}+\mathrm{h.c.}\right]\, .
\label{eqn:ac}
\end{align}
In Eq.~(\ref{eqn:ac}) the label $\sigma$ refers to the eigenstate $|\sigma\rangle$ of the internal plus static field Hamiltonian $\hat{H}_{\mathrm{in}}+\hat{H}_{\mathrm{F}}$, $d_{\sigma\sigma'}\equiv \langle \sigma |d_1|\sigma'\rangle$, and $E_{\sigma}$ is the energy of state $|\sigma\rangle$.  

Assembling all the many-body terms expressed in this basis, we obtain the time-dependent Hamiltonian
\begin{align}
\nonumber\hat{H}&=\sum_{\sigma}E_{\sigma}\sum_i\hat{n}_{i\sigma}-\sum_{\sigma}t_{\sigma}\sum_{\langle i,j\rangle}\left[\hat{a}_{i\sigma}^{\dagger}\hat{a}_{j\sigma}+\mathrm{h.c.}\right]\\
\nonumber&+\frac{1}{2}\sum_{\sigma,\sigma'}U_{\sigma\sigma'}\sum_{\langle i,j\rangle}\hat{n}_{i\sigma}\hat{n}_{j\sigma'}\\
&-\frac{1}{2}\sum_{\sigma\sigma'}d_{\sigma\sigma'}E_{\mathrm{AC}}\sum_i\left[\hat{a}_{i\sigma}^{\dagger}\hat{a}_{i\sigma'}e^{i\omega t}+\mathrm{h.c.}\right]\, .
\end{align}
If we change to a basis where all single-molecule states with $N=1$ rotate with frequency $\omega$ we have, finally:
\begin{align}
\nonumber\hat{H}&=\sum_{\sigma}\Delta_{\sigma}\sum_i\hat{n}_{i\sigma}-\sum_{\sigma}t_{\sigma}\sum_{\langle i,j\rangle}\left[\hat{a}_{i\sigma}^{\dagger}\hat{a}_{j\sigma}+\mathrm{h.c.}\right]\\
\nonumber&+\frac{1}{2}\sum_{\sigma,\sigma'}U_{\sigma\sigma'}\sum_{\langle i,j\rangle}\hat{n}_{i\sigma}\hat{n}_{j\sigma'}\\
&-\frac{1}{2}\sum_{\sigma\sigma'}d_{\sigma\sigma'}E_{\mathrm{AC}}\sum_i\left[\hat{a}_{i\sigma}^{\dagger}\hat{a}_{i\sigma'}+\mathrm{h.c.}\right]\, ,
\end{align}
where $\Delta_{\sigma}=E_{\sigma}$ for states with $N=0$ and $E_{\sigma}-\omega$ for states with $N=1$.

\section{Conclusions}
\label{sec:conclusion}
We have presented and derived the Hyperfine Molecular Hubbard Hamiltonian (HMHH). The HMHH is a lattice Hamiltonian describing the effective low-energy physics of an ultracold gas of heteronuclear alkali dimer molecules with hyperfine structure loaded into a 1D optical lattice and interacting with external DC electric, AC microwave, and static magnetic fields.  By tuning the angle between the electric and magnetic fields and the strength of the magnetic and AC fields it is possible to change the number and timescale of internal states contributing to many-body dynamics.  The Hamiltonian also displays emergent quantum dephasing, and has a phase diagram which depends strongly on the initial state.  These features make the HMHH an ideal candidate for a model quantum complex system.

Future work will involve time-evolving block decimation simulations of the HMHH similar to past studies of molecular Hubbard Hamiltonians~\cite{carr2009a}.  In particular, we will discuss measures of complexity and how they relate to experimentally measurable quantities.  Future work on the Hamiltonian itself will include realistic models of molecule loss due to inelastic and chemical processes.  Such dissipative processes are key to dissipative quantum phase transitions, which is a major area of interest in quantum many-body theory~\cite{chakravarty2010, lehur2010, Werner2010}.

We acknowledge useful discussions with Immanuel Bloch, John Bohn, Silke Ospelkaus, Luis Santos, and Peter Zoller.  This work was supported by the National Science Foundation under Grant PHY-0903457.

\appendix
\section{The Internal Hamiltonian}
\label{sec:internal}
A $^1\Sigma$ molecule in its electronic and vibrational ground states has three angular momentum degrees of freedom: the rotational angular momentum $\mathbf{N}$ and the nuclear spins $\mathbf{I}_1$ and $\mathbf{I}_2$.  In this work we shall use the coupling schemes $|\left(I_1I_2\right)INFM_F\rangle$ and $|I_1M_1I_2M_2NM_N\rangle$, which we refer to as the coupled and uncoupled bases, respectively.  Explicit expressions for all single-molecule matrix elements in both bases are provided in Appendix \ref{sec:matelem}.  The relevant Hamiltonian for the internal degrees of freedom $\hat{H}_{\mathrm{in}}$ may be written as a sum of rotational and hyperfine terms as
\begin{align}
\label{eq:internal}\hat{H}_{\mathrm{in}}=\hat{H}_{\mathrm{rot}}+\hat{H}_{\mathrm{hf}}
\end{align}
where
\begin{align}
\label{eq:rot}\hat{H}_{\mathrm{rot}}&=B_N\mathbf{N}^2\, ,\\
\label{eq:hf}\hat{H}_{\mathrm{hf}}&=\sum_{i=1}^{2}c_i\mathbf{N}\cdot\mathbf{I}_i+c_3\mathbf{I}_1\cdot\tilde{T}\cdot\mathbf{I}_2+c_4\mathbf{I}_1\cdot\mathbf{I}_2+\sum_{i=1}^2\mathbf{V}_i\cdot\mathbf{Q}_i\, .
\end{align}
The rotational term Eq.~\eqref{eq:rot} corresponds to the Hamiltonian of a rigid spherical rotor with $(2N+1)$-fold degenerate eigenstates $|NM_N\rangle$, $M_N$ being the projection of $\mathbf{N}$ on a space-fixed quantization axis~\cite{RSODM}.  The eigenenergies are given by $E_{NM_N}=B_NN\left(N+1\right)$, where $B_N$ is the rotational constant of the molecule (we use the notation $B_N$ instead of the more common $B$ to avoid confusion with the magnetic field magnitude $B$).  In the case of $^{40}$K$^{87}$Rb,  $B_N$=1.114 GHz~\cite{Hutson2008}.  The rotational level splitting defines the largest intrinsic energy scale for $^1\Sigma$ molecules.

The first term of the hyperfine Hamiltonian, $\sum_{i=1}^2c_i\mathbf{N}\cdot\mathbf{I}_i$ represents the interaction of the nuclear spins with the magnetic field created by the rotation of the molecule, and is governed by two coupling constants $c_{\mathrm{K}}$ and $c_{\mathrm{Rb}}$ related to the nuclear shielding tensor.  For $^{40}$K$^{87}$Rb, these have been determined from density functional calculations to be $\sim$20Hz and $\sim$100Hz, respectively~\cite{Hutson2008}.  Because of the smallness of these constants and the fact that this term does not couple states with different $N$, this term plays a very small role in the spectra.

The two nuclear spins have nuclear magnetic moments which interact via a resonant dipole-dipole interaction
\begin{align}
\hat{H}_{\mathrm{hf-dd}}&=g_H^2\mu_N^2\left(\mu_0/4\pi\right)\left[\frac{\mathbf{I}_1\cdot\mathbf{I}_2}{R^3}-\frac{3\left(\mathbf{I}_1\cdot\mathbf{R}\right)\left(\mathbf{R}\cdot\mathbf{I}_2\right)}{R^5}\right]\, ,
\end{align}
where $g_H$ is the proton $g$-factor and $\mathbf{R}$ the vector joining the two nuclei~\cite{RSODM}.  This may be written as the contraction of two rank-2 spherical tensors as
\begin{align}
\hat{H}_ {\mathrm{hf-dd}}&=-g_H^2\mu_N^2\left(\mu_0/4\pi\right)\langle R^{-3}\rangle\sqrt{6}\left(\mathbf{C}\right)^{\left(2\right)}\cdot \left(\mathbf{T}\left(\mathbf{I}_1,\mathbf{I}_2\right)\right)^{\left(2\right)}\,
\end{align}
where $\left(\mathbf{C}\right)^{\left(2\right)}$ is an unnormalized spherical harmonic in the relative degrees of freedom.  The nuclear spins can also interact indirectly through the electron spins, and do so even for $^1\Sigma$ configurations~\cite{RSODM}.  This indirect interaction is represented by a tensor $\tilde{J}$ which may be decomposed into its isotropic part $J_{\mathrm{iso}}$ and its anisotropy $\Delta J=J_{\parallel}-J_{\perp}$.  The combination of direct and indirect nuclear spin-nuclear spin interaction may thus be written as the sum of a scalar interaction and a tensor interaction as
\begin{align}
\hat{H}_{\mathrm{hf-dd}}+\hat{H}_{\mathrm{indirect}}&=c_4\mathbf{I}_1\cdot\mathbf{I}_2+c_3\mathbf{I}_1\cdot\tilde{T}\cdot \mathbf{I}_2
\end{align}
where $c_4\equiv J_{\mathrm{iso}}$, $c_3\equiv g_H^2\mu_N^2\left(\mu_0/4\pi\right)\langle R^{-3}\rangle-{\Delta J}/{3}$, and the tensor $\tilde{T}$ contains the angular dependence of the tensor interaction.  $c_3$ is of order 10Hz for the various isotopes of KRb, and so plays a very small role in the spectra.  $c_4$ splits the various levels according to their total nuclear spin $I$ as
\begin{align}
\nonumber&\langle \left(I_1I_2\right)INFM_F|c_4\mathbf{I}_1\cdot\mathbf{I}_2|\left(I_1I_2\right)I'N'F'M_F'\rangle\\
\nonumber &=\delta_{I,I'}\delta_{N,N'}\delta_{F,F'}\delta_{M_F,M_F'}\\
&\times\frac{c_4}{2}\left[I\left(I+1\right)-I_1\left(I_1+1\right)-I_2\left(I_2+1\right)\right]\, .
\end{align}
$c_4$ is of order 100Hz-10kHz for isotopes of KRb, and so is the dominant hyperfine contribution for $N=0$ in the absence of external fields, see Fig.~\ref{fig:kHZStark}.  Note that $c_4$ may be either positive or negative.  For $^{40}$K$^{87}$Rb, $c_4$=-20.304kHz~\cite{Hutson2008}, and so the lowest energy states for $N=0$ in zero field are the highest nuclear spin states $I=11/2$.

The final term in the hyperfine Hamiltonian is the interaction of the quadrupole moment of the nuclei with the gradient of the electric field produced by the electrons.  We may represent this interaction by the sum $\sum_{i=1}^{2}\mathbf{V}_i\cdot\mathbf{Q}_i$ where $\mathbf{V}_i$ is a second rank spherical tensor describing the electric field gradient at the $i^{th}$ nucleus and $\mathbf{Q}_i$ is a second rank spherical tensor describing the nuclear quadrupole of the $i^{th}$ nucleus.  The pertinent coupling constants $\left(eqQ\right)_i$ which arise in the matrix elements of this Hamiltonian are of order 100-1000kHz, making it the largest term in the hyperfine Hamiltonian.  The quadrupole term doesn't affect the $N=0$ level, however, and so the scalar spin-spin coupling dominates there.  In a strong DC field the rotational levels become deeply mixed and the nuclear quadrupole thus becomes the dominant hyperfine contribution for all states.

\section{Interactions with static external fields}
\label{sec:fields}
Polar molecules such as heteronuclear dimers can couple to external fields either through their permanent electric dipole moment, through magnetic moments generated from their rotation or nuclear spin, or through their polarizability tensor.  The Hamiltonian representing interaction of the molecule with a static DC electric field $\mathbf{E}_{\mathrm{DC}}$ and a static magnetic field $\mathbf{B}$ may be written
\begin{align}
\hat{H}_{\mathrm{F}}=&-\mathbf{d}\cdot\mathbf{E}_{\mathrm{DC}}-g_r\mu_N\mathbf{N}\cdot\mathbf{B}-\sum_{i=1}^{2}g_i\mu_N\mathbf{I}_i\cdot\mathbf{B}\left(1-\sigma_i\right)\, .
\end{align}
For $^1\Sigma$ molecules the permanent dipole moment $\mathbf{d}$ lies along the internuclear axis which defines the $p=0$ axis in a spherical coordinate system rotating with the molecule.  Because this basis leads to anomalous commutation relations $\left[J_i,J_k\right]=-i\hbar\epsilon_{ijk}J_k$~\cite{vanvleck51} we find it convenient to transform to the space-fixed frame where the angular momentum operators satisfy the normal commutation relations $\left[J_i,J_k\right]=i\hbar\epsilon_{ijk}J_k$, giving $\mathbf{d}\cdot \mathbf{e}_q\equiv d_q=d C^{\left(1\right)}_q\left(\theta,\phi\right)$, where $\mathbf{e}_q$ is a unit vector along the space fixed spherical $q$ direction and $C^{\left(1\right)}_q\left(\theta,\phi\right)$ is an unnormalized spherical harmonic whose arguments $\theta$ and $\phi$ are the polar and azimuthal angles of the internuclear axis in the space fixed frame.  Taking matrix elements of $d_q$ in our two basis sets yields
\begin{align}
\nonumber&\langle I_1M_1I_2M_2NM_N|d_q|I_1M_1'I_2M_2'N'M_N'\rangle\\
\nonumber&=\delta_{M_1,M_1'}\delta_{M_2,M_2'}d\sqrt{\left(2N+1\right)\left(2N'+1\right)}\left(-1\right)^{M_N}\\
&\times\left(\begin{array}{ccc} N&1&N'\\0&0&0\end{array}\right)\left(\begin{array}{ccc} N&1&N'\\-M_N&q&M_N'\end{array}\right)\, ,
\end{align}
\begin{align}
\nonumber&\langle\left(I_1I_2\right)INFM_F|d_q|\left(I_1I_2\right)I'N'F'M_F'\rangle\\
\nonumber&=\delta_{I,I'}d\left(-1\right)^{2F-M_F+I+N'+N+1}\left(\begin{array}{ccc} N&1&N'\\0&0&0\end{array}\right)\\
\nonumber&\times\sqrt{\left(2N+1\right)\left(2N'+1\right)\left(2F+1\right)\left(2F'+1\right)}\\
&\times\left(\begin{array}{ccc} N&1&N'\\-M_N&q&M_N'\end{array}\right)\left\{\begin{array}{ccc} N&F&I\\ F'&N'&1\end{array}\right\}\, ,
\end{align}
where $\left(\begin{array}{ccc} j_1&j_2&j_3\\ m_1&m_2&m_3\end{array}\right)$ is a Wigner 3-$j$ coefficient and $\left\{\begin{array}{ccc} j_1&j_2&j_3\\ j_4&j_5&j_6\end{array}\right\}$ is a Wigner 6-$j$ coefficient~\cite{Zare}.  We see that the rotational eigenstates have no net dipole moment, but that the dipole operator couples the state $| N,F,M_F\rangle$ with the states $| N\pm 1,F\pm 1, M_F+q\rangle$.  The introduction of a DC electric field $\mathbf{E}_{\mathrm{DC}}$ with Hamiltonian $-\mathbf{d}\cdot\mathbf{E}_{\mathrm{DC}}$ couples these levels and induces dipole moments, breaking the rotational symmetry and removing the $\left(2N+1\right)$-fold degeneracy.  Typical molecular dipole moments are measured in Debye (D), where 1D=503.4MHz/(kV/cm), and so the DC field becomes the dominant contribution to the Hamiltonian for modest fields of a few kV/cm.  The permanent dipole moment of KRb has been experimentally determined to be 0.566D~\cite{Ospelkaus2009b}.

%%%%%%%%%%% figure 1 %%%%%%%%%%%
%
\begin{figure}[t]
\begin{center}
\epsfxsize=9cm \epsfysize=6.2 cm \leavevmode \epsfbox{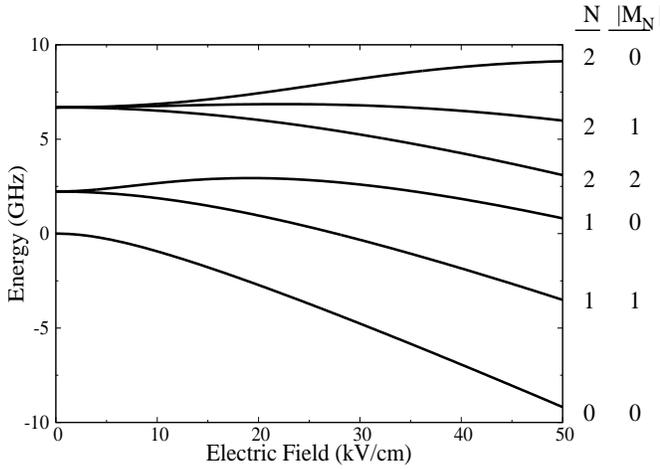}
\caption{\label{fig:GHZStark}  \emph{GHz scale view of the Stark effect for KRb}.  Introduction of a DC field breaks the degeneracy between all states with the same $N$ but different $\left|M_N\right|$.  The large electric dipole moment causes GHz scale energy shifts which completely obscure the hyperfine splittings on the scale of this plot.  Because the dipole moment is the same for any isotope of KRb, the Stark effect on this scale is the same for all isotopes.}
\end{center}
\end{figure}

On the scale of the rotational constant, the effect of a DC field on the single-molecule energy spectrum is as in Fig.~\ref{fig:GHZStark}.  It is quadratic for field energies small compared to the rotational energy but becomes linear in stronger fields because the field strongly mixes states of opposite parity~\cite{townes}.  We consider the quantization axis to lie along the field direction, and so states with the same value of $\left|M_N\right|$ remain degenerate.  A universal plot for all $^1\Sigma$ molecules results on this scale if the energy and field strength $dE_{\mathrm{DC}}$ are both scaled to the rotational constant.

%%%%%%%%%%% figure 1 %%%%%%%%%%%
%
\begin{figure}[t]
\begin{center}
\epsfxsize=9cm \epsfysize=6.2 cm \leavevmode \epsfbox{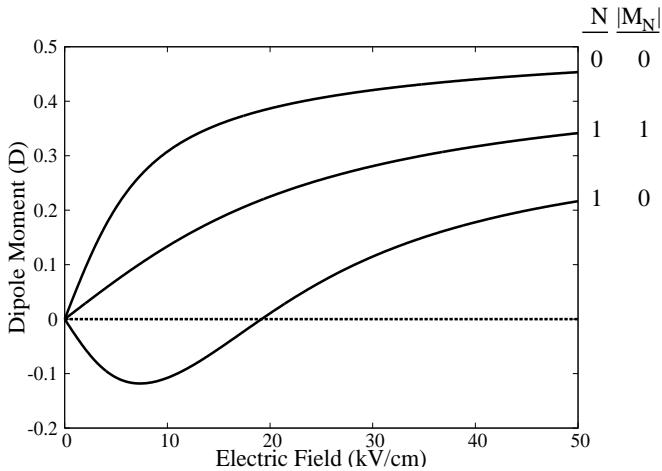}
\caption{\label{fig:indDip}  \emph{Induced dipoles for KRb in an electric field}.  The $N=0$ and $N=1$, $M_N=\pm1$ levels orient along the field, giving rise to positive dipole moments.  The $N=1$, $M_N=0$ state antialigns with the field for small fields, but aligns in stronger fields.  All resonant dipole moments approach the ``permanent'' value 0.566D as the field strength increases.}
\end{center}
\end{figure}

The average orientation of the molecule with the electric field can be obtained with the Feynman-Hellman theorem as
\begin{align}
 \langle \cos\theta\rangle&=-\frac{\partial E}{\partial \left(dE_{\mathrm{DC}}\right)}\, ,
\end{align}
where $E$ is the energy eigenvalue.  The energy eigenvalue is dominated by the GHz scale structure, thus the degree of alignment with the field is essentially independent of the hyperfine structure.  From the degree of orientation we can also determine the effective space-fixed dipole moment as $d\langle\cos\theta\rangle$.  Fig.~\ref{fig:indDip} shows the behavior of the induced dipoles as the field strength is increased.  For all field strengths the $N=0$ and $N=1$, $M_N=\pm1$ states align with the field and so have a positive induced dipole moment.  In contrast, the $N=1$, $M_N=0$ state antialigns with the field for weak fields and aligns with the field for stronger fields.

%%%%%%%%%%% figure 2 %%%%%%%%%%%
%
\begin{figure}[t]
\begin{center}
\epsfxsize=9cm \epsfysize=6.2 cm \leavevmode \epsfbox{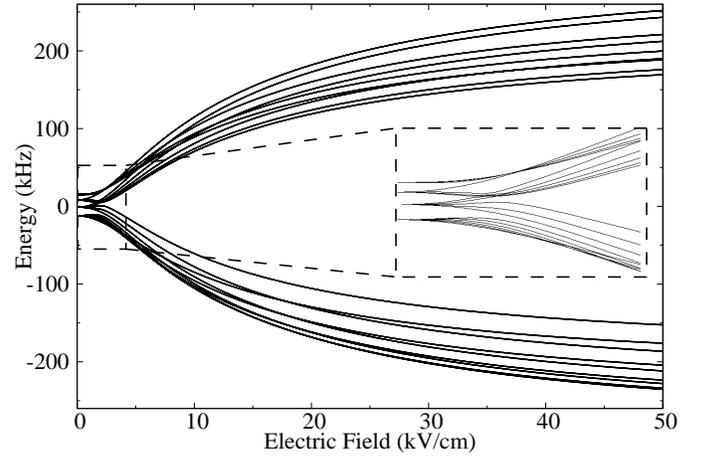}
\caption{\label{fig:kHZStark}  \emph{kHz scale view of the Stark effect for $^{40}$K$^{87}$Rb, $N$=0}.  All energies are shown relative to the GHz scale field-dependent average energy for $N=0$, see Fig.~\ref{fig:GHZStark}.  The inset shows the weak field region where the scalar spin-spin interaction has split the levels according to $I$ (equivalently, $F$), with larger $I$ having lower energy.  As the field is increased the nuclear quadrupole couplings split according to $M_I$, and in large fields $M_{\mathrm{Rb}}$ and $M_{\mathrm{K}}$ also become well defined.  See text for details.}
\end{center}
\end{figure}
The magnitude of the field energy completely obscures the hyperfine splittings, and so to see the effects of hyperfine structure we subtract from each state with a given $N$ the field-dependent average energy of all hyperfine states with the same $N$.  For $N=0$ the results are shown in Fig.~\ref{fig:kHZStark}.   For low fields the hyperfine splittings are dominated by the scalar spin-spin coupling and are of order $c_4$, a few kHz.  As the field is increased the various hyperfine states split according to $\left|M_I\right|$.  For large fields $M_{\mathrm{1}}$ and $M_{\mathrm{2}}$ also become well defined, which occurs because the energetic differences between states with $\Delta M_N=\pm1$ become larger than the quadrupole coupling constants (see Eq.~\eqref{eq:quaduc}).  Pairs of $M_1$ and $M_2$ which have the same $\left|M_1+M_2\right|$ are degenerate, and the state with $\left|M_1+M_2\right|=0$ is degenerate due to reflection symmetry in the plane of the electric field vector.

Because of the signs of the quadrupole couplings for $^{40}$K$^{87}$Rb, the lowest energy states are those with $M_{\mathrm{Rb}}$ the largest and $M_{\mathrm{K}}$ the smallest.  Because the kHz scale Stark effect depends on several molecular parameters it cannot be put into a universal form for all $^1\Sigma$ molecules like the GHz scale Stark effect.  However, the qualitative structure will be similar for all $^1\Sigma$ molecules with nuclear quadrupole couplings; key differences being the number of nondegenerate levels and the energetic ordering of the magnetic quantum numbers~\cite{Hutson2008}.  The hyperfine Stark effect for $N=1$ and other molecular species as well as the effects of electric fields on microwave spectra may be found in Ref.~\cite{Ran2009}.

Magnetic fields couple to the magnetic moments generated by the rotation of the molecule and by the nuclear spins.  The former interaction is given by $-g_r\mu_N\mathbf{N}\cdot\mathbf{B}$, where $g_r$ is the rotational $g$ factor of the molecule and $\mu_N$ is the nuclear magneton $e\hbar/2 m_p$=762.259Hz/G~\cite{CODATA}.  The latter interaction is given by $-\sum_{i=1}^{2}g_i\mu_N\mathbf{I}_i\cdot\mathbf{B}\left(1-\sigma_i\right)$, where $g_1$ and $g_2$ are the g-factors of nucleus 1 and 2, respectively, and $\sigma_i$ is the isotropic part of the nuclear shielding tensor for nucleus $i$.  The rotational contribution is typically much smaller than the contributions from the nuclei, due to smaller g-factors and the fact that the isotropic parts of the nuclear shielding tensors are typically only a few parts per thousand.  For example, in $^{40}$K$^{87}$Rb $g_r=0.0140$, $g_{\mathrm{K}}=-0.324$, $g_{\mathrm{Rb}}=1.834$, $\sigma_{\mathrm{K}}=1321$ppm, and $\sigma_{\mathrm{Rb}}=3469$ppm~\cite{Hutson2008}.  We neglect diamagnetic contributions to the Zeeman effect, as these contributions are small for the fields we consider.

%%%%%%%%%%% figure 3 %%%%%%%%%%%
%
\begin{figure}[t]
\begin{center}
\epsfxsize=9cm \epsfysize=6.2 cm \leavevmode \epsfbox{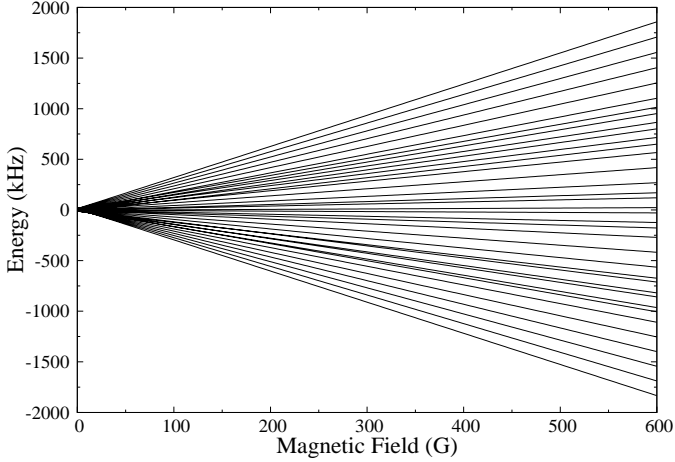}
\caption{\label{fig:N0Zeeman}  \emph{Zeeman effect for $^{40}$K$^{87}$Rb, $N$=0}. The magnetic field  splits the hyperfine levels according to their projections $M_{\mathrm{K}}$ and $M_{\mathrm{Rb}}$ with splittings between adjacent levels of order kHz for the experimentally relevant range $B\sim 550$G.  The lowest (highest) energy state corresponds to $m_F=-4+3/2=-5/2$ ($5/2$).  The zero field splitting is set by $c_4$ and is not visible on the scale of this plot.}
\end{center}
\end{figure}

%%%%%%%%%%% figure ??? %%%%%%%%%%%
%
\begin{figure}[t]
\begin{center}
\epsfxsize=9cm \epsfysize=6.2 cm \leavevmode \epsfbox{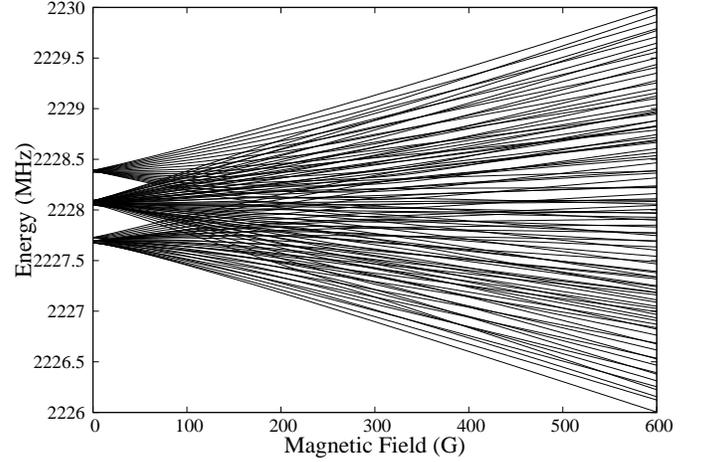}
\caption{\label{fig:N1Zeeman}  \emph{Zeeman effect for $^{40}$K$^{87}$Rb, $N$=1}. The zero field splitting is caused mainly by the nuclear quadrupole intraction and separates the levels into groups of well defined $F$.  The much larger zero field splitting causes avoided crossings between states with the same $M_F$ to occur at much higher fields than in the $N=0$ case.}
\end{center}
\end{figure}

Typical experimental magnetic fields are $\sim550$G because of the Feshbach association stage of the STIRAP procedure~\cite{niKK2008}.  In Fig.~\ref{fig:N0Zeeman} we show the Zeeman effect for the $N=0$ level of $^{40}$K$^{87}$Rb for fields up to this range.  We see that the magnetic field splits the spectrum according to the nuclear spin projections $M_{\mathrm{K}}$ and $M_{\mathrm{Rb}}$, with larger (smaller) $M_{\mathrm{K}}$ ($M_{\mathrm{Rb}}$) having lower energy due to the signs of the $g$-factors for $^{40}$K$^{87}$Rb.  Because the nuclear quadrupole interaction doesn't affect the $N=0$ level, the zero field splittings are determined by the small scalar spin-spin coupling parameter $c_4$.  The Zeeman term dominates over the scalar spin-spin coupling at very low fields and so the effects of the scalar spin-spin coupling are not discernible on the scale of this plot.  Additionally, the Zeeman contribution at these fields is larger than the hyperfine Stark splittings from the largest electric fields accessible in current experiments, see Fig.~\ref{fig:kHZStark}.

The spectrum for the $N=1$ level of $^{40}$K$^{87}$Rb is shown in Fig.~\ref{fig:N0Zeeman}.  It is greatly complicated by the fact that there are three times as many states as the $N=0$ case (corresponding to the allowed $M_N$).  Also, the nuclear quadrupole interaction affects the $N=1$ level, causing the large zero field splittings.  These larger zero field splittings delay the separation of the levels into well defined $M_1$ and $M_2$, and also causes a complicated series of avoided crossings between states with the same $M_F$.

\section{Explicit values for the single-molecule matrix elements}
\label{sec:matelem}
Here we present the matrix elements of the single-molecule terms of the Hamiltonian \eqref{eq:internal} in the coupled and uncoupled basis sets.  We adopt the conventions of Zare~\cite{Zare}.

The matrix elements of the rotational Hamiltonian are given by
 \begin{align}
\nonumber&\langle I_1M_1I_2M_2NM_N|B_N\mathbf{N}^2|I_1M_1'I_2M_2'N'M_N'\rangle\\
&=\delta_{M_1,M_1'}\delta_{M_2,M_2'}\delta_{N,N'}\delta_{M_N,M_N'}B_N N\left(N+1\right)\, ,
\end{align}
\begin{align}
\nonumber&\langle \left(I_1I_2\right)INFM_F|B_N\mathbf{N}^2| \left(I_1I_2\right)I'N'F'M_F'\rangle\\
&=\delta_{I,I'}\delta_{N,N'}\delta_{F,F'}\delta_{M_F,M_F'}B_N N\left(N+1\right)\, .
\end{align}

The matrix elements of the rotation-spin Hamiltonian are given by
\begin{align}
\nonumber&\langle I_1M_1I_2M_2NM_N|\sum_{i=1}^{2}c_i\mathbf{N}\cdot\mathbf{I}_i|I_1M_1'I_2M_2'N'M_N'\rangle\\
\nonumber&=\delta_{N,N'}\sum_{q}\left(-1\right)^{q+N-M_N}\left(\begin{array}{ccc} N&1&N\\ -M_N&q&M_N'\end{array}\right)\\
\nonumber&\times \sum_{i}c_i\left(-1\right)^{I_i-M_i}\left(\begin{array}{ccc} I_i&1&I_i\\ -M_i&-q&M_i'\end{array}\right)\\
&\times \sqrt{N\left(2N+1\right)\left(N+1\right)I_i\left(2I_i+1\right)\left(I_i+1\right)}\, ,
\end{align}
\begin{align}
\nonumber&\langle \left(I_1I_2\right)INFM_F|\sum_{i=1}^{2}c_i\mathbf{N}\cdot\mathbf{I}_i| \left(I_1I_2\right)I'N'F'M_F'\rangle\\
\nonumber&=\delta_{N,N'}\left(-1\right)^{I+N+F+I_1+I_2+1}\left\{\begin{array}{ccc} I&N&F\\ N&I'&1\end{array}\right\}\\
\nonumber&\times\sqrt{N\left(2N+1\right)\left(N+1\right)\left(I+1\right)\left(2I+1\right)\left(2I'+1\right)}\\
\nonumber &\times\Bigg[\delta_{M_2,M_2'}\left(-1\right)^{I'}c_1\left\{\begin{array}{ccc} I_1&I&I_2\\ I'&I_1&1\end{array}\right\}\sqrt{I_1\left(2I_1+1\right)\left(I_1+1\right)}\\
&+\delta_{M_1,M_1'}\left(-1\right)^{I}c_2\left\{\begin{array}{ccc} I_2&I&I_1\\ I'&I_2&1\end{array}\right\}\sqrt{I_2\left(2I_2+1\right)\left(I_2+1\right)}\Bigg]\, .
\end{align}

The matrix elements of the scalar spin-spin coupling are
\begin{align}
\nonumber&\langle I_1M_1I_2M_2NM_N|c_4\mathbf{I}_1\cdot\mathbf{I}_2|I_1M_1'I_2M_2'N'M_N'\rangle\\
\nonumber&=\delta_{N,N'}\delta_{F,F'}\delta_{M_F,M_F'}c_4\left(-1\right)^{I_1-M_1+I_2-M_2}\\
\nonumber&\times\sqrt{\left(2I_1+1\right)I_1\left(I_1+1\right)\left(2I_2+1\right)I_2\left(I_2+1\right)}\\
&\times\sum_q\left(-1\right)^{q}\left(\begin{array}{ccc} I_1&1&I_1\\ -M_1&q&M_1'\end{array}\right)\left(\begin{array}{ccc} I_2&1&I_2\\ -M_2&-q&M_2'\end{array}\right)\, ,
\end{align}
\begin{align}
\nonumber&\langle \left(I_1I_2\right)INFM_F|c_4\mathbf{I}_1\cdot\mathbf{I}_2|\left(I_1I_2\right)I'N'F'M_F'\rangle\\
\nonumber &=\delta_{I,I'}\delta_{N,N'}\delta_{F,F'}\delta_{M_F,M_F'}\\
&\times\frac{c_4}{2}\left[I\left(I+1\right)-I_1\left(I_1+1\right)-I_2\left(I_2+1\right)\right]\, .
\end{align}

The matrix elements of the tensor spin-spin coupling are
\begin{align}
\nonumber&\langle I_1M_1I_2M_2NM_N|c_3\mathbf{I}_1\cdot\tilde{T}\cdot\mathbf{I}_2|I_1M_1'I_2M_2'N'M_N'\rangle\\
\nonumber&=-c_3\sqrt{6}\left(\begin{array}{ccc} N&2&N'\\ 0&0&0\end{array}\right)\sqrt{\left(2N+1\right)\left(2N'+1\right)}\\
\nonumber &\times \sqrt{I_1I_2\left(2I_1+1\right)\left(2I_2+1\right)\left(I_1+1\right)\left(I_2+1\right)}\\
\nonumber&\times\sum_q\left(-1\right)^{q-M_N+I_1-M_1+I_2-M_2}\left(\begin{array}{ccc} N&2&N'\\ -M_N&q&M_N'\end{array}\right)\\
\nonumber&\times \sum_m\langle 1,m; 1,-q-m|2,-q\rangle \left(\begin{array}{ccc} I_1&1&I_1\\ -M_1&m&M_1'\end{array}\right)\\
&\times \left(\begin{array}{ccc} I_2&1&I_2\\ -M_2&-q-m&M_2'\end{array}\right)\, ,
\end{align}
\begin{align}
\nonumber&\langle \left(I_1I_2\right)INFM_F|c_3\mathbf{I}_1\cdot\tilde{T}\cdot\mathbf{I}_2|\left(I_1I_2\right)I'N'F'M_F'\rangle\\
\nonumber &=-c_3\delta_{F,F'}\delta_{M_F,M_F'}\left(-1\right)^{I'+F}\left\{\begin{array}{ccc} I&N&F\\ N'&I'&2\end{array}\right\}\\
\nonumber&\times\sqrt{\left(2N+1\right)\left(2N'+1\right)}\left(\begin{array}{ccc} N&2&N'\\ 0&0&0\end{array}\right)\left\{\begin{array}{ccc} I_1&I_1&1\\ I_2&I_2&1\\ I&I'&2\end{array}\right\}\\
\nonumber&\times \sqrt{30\left(2I+1\right)\left(2I'+1\right)I_1I_2}\\
&\times \sqrt{\left(I_1+1\right)\left(I_2+1\right)\left(2I_1+1\right)\left(2I_2+1\right)} \, .
\end{align}

The matrix elements of the nuclear quadrupole Hamiltonian are given by
\begin{align}
\label{eq:quaduc}\nonumber&\langle I_1M_1I_2M_2NM_N|\sum_{i=1}^{2}\mathbf{V}_i\cdot\mathbf{Q}_i|I_1M_1'I_2M_2'N'M_N'\rangle\\
\nonumber&=\sum_{i=1}^{2}\frac{\left(eqQ\right)_i}{4}\sum_{q}\left(-1\right)^{q-M_N+I_i-M_i}\sqrt{\left(2N+1\right)\left(2N'+1\right)}\\
\nonumber&\times \left(\begin{array}{ccc} N&2&N'\\ -M_N&q&M_N'\end{array}\right)\left(\begin{array}{ccc} I_i&2&I_i\\ -M_i&-q&M_i'\end{array}\right)\\
&\times \left(\begin{array}{ccc} N&2&N'\\ 0&0&0\end{array}\right)\left(\begin{array}{ccc} I_i&2&I_i\\ -I_i&0&I_i\end{array}\right)^{-1}\, ,
\end{align}
\begin{align}
\nonumber&\langle \left(I_1I_2\right)INFM_F|\sum_{i=1}^{2}\mathbf{V}_i\cdot\mathbf{Q}_i| \left(I_1I_2\right)I'N'F'M_F'\rangle\\
\nonumber&=\delta_{F,F'}\delta_{M_F,M_F'}\frac{1}{4}\left(-1\right)^{I'+F+I_1+I_2}\\
\nonumber&\times\sqrt{\left(2N+1\right)\left(2N'+1\right)\left(2I'+1\right)\left(2I+1\right)}\\
\nonumber&\times\left(\begin{array}{ccc} N&2&N'\\ 0&0&0\end{array}\right)\left\{\begin{array}{ccc} I&N&F\\ N'&I'&2\end{array}\right\}\\
\nonumber &\times \Big[\delta_{I_2,I_2'}\left(eqQ\right)_1\left(-1\right)^{I'}\left\{\begin{array}{ccc} I_1&I&I_2\\ I'&I_1&2\end{array}\right\} \left(\begin{array}{ccc} I_1&2&I_1\\ -I_1&0&I_1\end{array}\right)^{-1}\\
&+\delta_{I_1,I_1'}\left(eqQ\right)_2\left(-1\right)^{I}\left\{\begin{array}{ccc} I_2&I&I_1\\ I'&I_2&2\end{array}\right\} \left(\begin{array}{ccc} I_2&2&I_2\\ -I_2&0&I_2\end{array}\right)^{-1}\Big]\, .
\end{align}

\bibliographystyle{prsty}
\bibliography{HMHH}

\begin{thebibliography}{10}

\bibitem{carr2009b}
L.~D. Carr, D. Demille, R.~V. Krems, and J. Ye, New J. Phys. {\bf 11},  055049
  (2009).

\bibitem{niKK2008}
K.-K. Ni, S. Ospelkaus, M.~H.~G. {de Miranda}, A. Pe{\'e}r, B. Neyenhuis, J.~J.
  Zirbel, S. Kotochigova, P.~S. Julienne, D.~S. Jin, and J. Ye, Science {\bf
  322},  231  (2008).

\bibitem{Ospelkaus2009b}
S. Ospelkaus, K.~K. Ni, M.~H.~G. {de Miranda}, A. Pe'er, B. Nyenhuis, D. Wang,
  S. Kotochigova, P.~S. Julienne, D.~S. Jin, , and J. Ye, Faraday Discuss. {\bf
  142},  351  (2009).

\bibitem{Deiglmayr2008b}
J. Deiglmayr, A. Grochola, M. Repp, K. M\"ortlbauer, C. Gl\"uck, J. Lange, O.
  Dulieu, R. Wester, and M. Weidem\"uller, Phys. Rev. Lett. {\bf 101},  133004
  (2008).

\bibitem{Danzl2010}
J.~G. Danzl, M.~J. Mark, E. Haller, M. Gustavsson, R. Hart, J. Aldegunde, J.~M.
  Hutson, and H.-C. N\"agerl, Nature Phys. {\bf 6},  265  (2010).

\bibitem{Winkler2007}
K. Winkler, F. Lang, G. Thalhammer, P.~v.~d. Straten, R. Grimm, and J.~H.
  Denschlag, Phys. Rev. Lett. {\bf 98},  043201  (2007).

\bibitem{Pilch2009}
K. Pilch, A.~D. Lange, A. Prantner, G. Kerner, F. Ferlaino, H.-C. N\"agerl, and
  R. Grimm, Phys. Rev. A {\bf 79},  042718  (2009).

\bibitem{Voigt2009}
A.-C. Voigt, M. Taglieber, L. Costa, T. Aoki, W. Wieser, T.~W. H\"ansch, and K.
  Dieckmann, Phys. Rev. Lett. {\bf 102},  020405  (2009).

\bibitem{Hutson2009}
J. Aldegunde, H. Ran, and J.~M. Hutson, Phys. Rev. A {\bf 80},  043410  (2009).

\bibitem{Ran2009}
H. Ran, J. Aldegunde, and J.~M. Hutson, e-print http://arxiv.org/abs/0909.3644
  (2009).

\bibitem{ospelkaus2009}
S. Ospelkaus, K.~K. Ni, G. Quemener, B. Neyenhuis, D. Wang, M.~H.~G. {de
  Miranda}, J.~L. Bohn, J. Ye, and D.~S. Jin, Phys. Rev. Lett. {\bf 104},
  030402  (2010).

\bibitem{feynmanRP1982}
R.~P. Feynman, Int. J. Theor. Phys. {\bf 21},  467  (1982).

\bibitem{lewensteinM2007}
M.~S.~A. Lewenstein, V. Ahufinger, B. Damski, A. {Sen De}, and U. Sen, Adv.
  Phys. {\bf 56},  243  (2007).

\bibitem{micheli2007}
A. Micheli, G. Pupillo, H.~P. B{\"u}chler, and P. Zoller, Phys. Rev. A {\bf
  76},  043604  (2007).

\bibitem{carr2009a}
M.~L. Wall and L.~D. Carr, New J. Phys. {\bf 11},  055027  (2009).

\bibitem{Brennen2007}
G.~K. Brennen, A. Micheli, and P. Zoller, New J. Phys. {\bf 9},  138  (2007).

\bibitem{huber2009}
S.~D. Huber and E. Altman, Phys. Rev. Lett. {\bf 103},  160402  (2009).

\bibitem{Barmettler2009}
P. Barmettler, M. Punk, V. Gritsev, E. Demler, and E. Altman, e-print
  arXiv:0911.1927  (2009).

\bibitem{A&M}
N. Ashcroft and D. Mermin, {\em Solid State Physics} (Saunders College
  Publishing, Orlando, 1976).

\bibitem{deiglmayr2008}
J. Deiglmayr, M. Aymar, R. Wester, M. Weidem{\"u}ller, and O. Dulieu, J. Chem.
  Phys. {\bf 129},  064309  (2008).

\bibitem{iskin2007b}
M. Iskin and C.~A.~R. {Sa de Melo}, Phys. Rev. A. {\bf 76},  013601  (2007).

\bibitem{wallforthcoming}
M.~L. Wall and L.~D. Carr, 2010, in preparation.

\bibitem{paredes2004}
B. Paredes, A. Widera, V. Murg, O. Mandel, S. Folling, I. Cirac, G.~V.
  Shlyapnikov, and T.~W. Hansch, Nature {\bf 429},  277  (2005).

\bibitem{tolra2004}
B.~L. Tolra, K.~M. O'Hara, J.~H. Huckans, W.~D. Phillips, S.~L. Rolston, and
  J.~V. Porto, Phys. Rev. Lett. {\bf 92},  190401  (2004).

\bibitem{kinoshita2006}
T.~W. {T. Kinoshita} and D.~S. Weiss, Nature {\bf 440},  900  (2006).

\bibitem{buechler2007}
H.~P. B{\"u}chler, A. Micheli, and P. Zoller, Nature Phys. {\bf 3},  726
  (2007).

\bibitem{Hutson2008}
J. Aldegunde, B.~A. Rivington, P.~S. \.{Z}uchowski, and J.~M. Hutson, Phys.
  Rev. A {\bf 78},  033434  (2008).

\bibitem{reyAM2004}
A.~M. Rey, Ph.D. thesis, University of Maryland, 2004.

\bibitem{niKK2010}
K.~K. Ni, S. Ospelkaus, D. Wang, G. Quemener, B. Neyenhuis, M.~H.~G. {de
  Miranda}, J.~L. Bohn, J. Ye, and D.~S. Jin, arXiv:1001.2809  (2010).

\bibitem{chakravarty2010}
S. Chakravarty, e-print http://arxiv.org/abs/0909.2316  (2010).

\bibitem{lehur2010}
K. {Le Hur}, e-print http://arxiv.org/abs/0909.4822  (2010).

\bibitem{Werner2010}
P. Werner and M. Troyer, Chapter of the book 'Understanding Quantum Phase
  Transitions', edited by Lincoln D. Carr (CRC Press / Taylor and Francis,
  2010)  (2010).

\bibitem{RSODM}
J. Brown and A. Carrington, {\em Rotational Spectroscopy of Diatomic Molecules}
  (Cambridge University Press, Cambridge, 2003).

\bibitem{vanvleck51}
J.~H.~V. Vleck, Rev. Mod. Phys. {\bf 23},  213  (1951).

\bibitem{Zare}
R. Zare, {\em Angular Momentum: Understanding Spatial Aspects in Chemistry and
  Physics} (Wiley, New York, 1988).

\bibitem{townes}
R.~N. Townes and A.~L. Schawlow, {\em Microwave Spectroscopy} (Dover, New York,
  1975).

\bibitem{CODATA}
P.~J. Mohr and B.~N. Taylor, Rev. Mod. Phys. {\bf 72},  351  (2000).

\end{thebibliography}

\end{document}